\RequirePackage{fix-cm} 
\documentclass[a4paper, twoside, reqno, 12pt, dvips]{amsart}
\usepackage{fixltx2e}   

\usepackage{eucal}
\usepackage{xspace}
\usepackage{amsgen}
\usepackage{amssymb}
\usepackage{amsmath}
\usepackage{amsfonts}
\usepackage{MnSymbol}
\usepackage[nice]{nicefrac}
\usepackage{mathrsfs, dsfont}
\usepackage{url}
\usepackage{a4wide}

\usepackage{acronym}

\usepackage{indentfirst}
\usepackage{graphicx}
\usepackage{psfrag}

\usepackage[usenames, dvipsnames, pdf]{pstricks}
\usepackage{epsfig}
\usepackage{pst-grad} 
\usepackage{pst-plot} 

\usepackage{subfigure}

\usepackage{longtable}

\usepackage[french,english]{babel}
\usepackage[latin1]{inputenc}

\usepackage{color}
\definecolor{oneblue}{rgb}{0,0.0,0.75}
\usepackage[colorlinks,
            urlcolor=oneblue,
            linkcolor=oneblue,
            citecolor=oneblue,
            bookmarksopen=false,
            pagebackref]{hyperref}

\numberwithin{equation}{section}

\newcommand{\depth}{d}

\newcommand{\ud}{\mathrm{d}}
\newcommand{\ui}{\mathrm{i}}
\newcommand{\ue}{\mathrm{e}}
\newcommand{\Ll}{\mathscr{L}}
\newcommand{\Nn}{\mathscr{N}}

\newcommand{\eps}{\varepsilon}
\renewcommand{\L}{\mathcal{L}}
\renewcommand{\O}{\mathcal{O}}
\newcommand{\Lf}{\mathfrak{L}}

\renewcommand{\Re}{\operatorname{Re}}
\renewcommand{\Im}{\operatorname{Im}}

\newcommand{\sur}[1]{{#1}_\text{s}}                    
\renewcommand{\bot}[1]{{#1}_\text{b}}                  

\newcommand{\sech}{\mathrm{sech}}

\newcommand{\half}{{\textstyle{1\over2}}}

\begin{document}

\title[Fast computation of solitary waves]%
{Fast accurate computation of the fully nonlinear solitary surface gravity waves}

\author[D. Clamond]{Didier Clamond$^*$}
\address{Laboratoire J.-A. Dieudonn\'e, Universit\'e de Nice -- Sophia Antipolis, 
Parc Valrose, 06108 Nice cedex 2, France}
\email{diderc@unice.fr}
\urladdr{\url{http://math.unice.fr/~didierc/}}
\thanks{$^*$ Corresponding author. Tel: +33 492 076 206, Fax: +33 493 517 974}

\author[D. Dutykh]{Denys Dutykh}
\address{University College Dublin, School of Mathematical Sciences, Belfield, Dublin 4, Ireland \and LAMA, UMR 5127 CNRS, Universit\'e de Savoie, Campus Scientifique, 73376 Le Bourget-du-Lac Cedex, France}
\email{Denys.Dutykh@ucd.ie}
\urladdr{\url{http://www.denys-dutykh.com/}}

\begin{abstract}
In this short note, we present an easy to implement and fast algorithm for the computation of the steady solitary gravity wave solution of the free surface Euler equations in irrotational motion. First, the problem is reformulated in a fixed domain using the conformal mapping technique. Second, the problem is reduced to a single equation for the free surface. Third, this equation is solved 
using Petviashvili's iterations together with pseudo-spectral discretisation. This method has a super-linear complexity, since the most demanding operations can be performed using a FFT algorithm. Moreover, when this algorithm is combined with the multi-precision floating point computations, the results can be obtained to any arbitrary accuracy.
\end{abstract}


\maketitle

\tableofcontents

\section{Introduction}

Solitary waves play a central role in nonlinear sciences. They appear in various fields ranging from plasmas physics to hydrodynamics and nonlinear optics \cite{infrow2000, Yang2010}. In some special cases, solitary wave solutions can be found analytically. For example, explicit expressions are known for integrable models such as KdV and NLS equations, but also for some non-integrable equations such as \cite{CD2012,Serre1953}. However, no closed form solutions are known for the practically very important case of the free surface Euler equations. In order to construct these solutions, one has to apply some approximation methods. Historically, asymptotic approximations have been proposed first \cite{Fenton1972} but, these series being divergent \cite{germain1967}, 
the high-order approximations are only valid in the limit $a \to 0$ ($a$ the wave amplitude). In order to avoid all these limitations, various numerical approaches have been also proposed, e.g. \cite{Fenton1982,Okasho2002}. We can mention the boundary integral equation method applied to the computation of solitary waves, see \cite{evansford96,Hunter1983,mak02,Parau2005} and the references therein. 

One of the most widely used methods nowadays is Tanaka's algorithm \cite{Tanaka1986}. In this note, we propose a simpler and faster method based on Petviashvili's iterations \cite{Petviashvili1976} leading to a very efficient numerical scheme for the computation of solitary gravity waves of the full Euler equations in the water of finite depth. Petviashvili's iterations have been successfully used for various simplified water wave equations but, to our knowledge, never for the fully nonlinear problem. The algorithm proposed here is efficient and it allows a very straightforward compact implementation under {\sc Matlab}, for example. Such a script can be freely downloaded \cite{Clamond2012}.

The note is organised as follows. In section \ref{sec:model}, we introduce the hypothesis, the notations and the Babenko equation which is equivalent to the original problem. In section \ref{sec:num}, we describe a very simple and fast algorithm to compute the solution, and we give some numerical results. Finally, the main conclusions are outlined in Section \ref{sec:concl}.

\section{Mathematical model}\label{sec:model}

We consider a steady two-dimensional potential flow due to a solitary gravity wave propagating at the surface of a homogeneous fluid layer of constant depth $d$. Let be $(x,y)$ a Cartesian coordinate system moving with the wave, $x$ being the horizontal coordinate and $y$ the upward vertical one. Since solitary waves are localised in space, the surface elevation tends to zero as $x \to \pm\infty$, and $x = 0$ is the abscissa of the crest. The equations of the bottom, of the free surface and of the mean water level are given correspondingly by $y = -\depth$, $y = \eta(x)$ and $y = 0$, respectively.

\subsection{Governing equation}\label{sec:bab}

Let be $\phi$, $\psi$, $u$ and $v$ the velocity potential, the stream function, the horizontal and vertical velocities, respectively, such that $u = \phi_x = \psi_y$ and $v = \phi_y = -\psi_x$. It is convenient to introduce the complex potential $f\equiv\phi+\ui\/\psi$ (with $\ui^2=-1$) and the complex velocity $w\equiv u-\ui\/v$ that are holomorphic functions of $z\equiv x+\ui\/y$ (i.e., $w = \ud f/\ud z$). The complex conjugate is denoted with a star (e.g., $z^\ast = x-\ui\/y$), while subscripts `b' denote the quantities written at the seabed --- e.g., $\bot{z}(x) = x-\ui\depth$, $\bot{\phi}(x)=\phi(x,y\!=\!-\depth)$ --- and subscripts `s' denote the quantities written at the surface --- e.g., $\sur{z}(x) = x+\ui\eta(x)$, $\sur{\phi}(x)=\phi(x,y\!=\!\eta(x))$. We emphasise that $\sur{\psi}$ and $\bot{\psi}$ are constants because the surface and the bottom are streamlines.

The far field velocity is such that  $(u,v)\to(-c,0)$ as $x\to\pm\infty$, so $c$ is the wave phase velocity observed in the frame of reference where the fluid is at rest at infinity ($c>0$ if the wave travels to the increasing $x$-direction). Note that $c = (\bot{\psi}-\sur{\psi})/\depth$ due to the mass conservation \cite{LH1974}.

The dynamic condition can be expressed in term of the Bernoulli equation
\begin{equation}\label{eq:bernbase}
2\,p\ +\ 2\,g\,y\ +\ u^2\ +\ v^2\ =\ c^2,
\end{equation}
where $p$ is the pressure divided by the density and $g > 0$ is the acceleration due to gravity. At the free surface the pressure equals that of the atmosphere which is constant and set to zero without loss of generality, i.e., $\sur{p} = 0$.

\subsection{Conformal mapping}

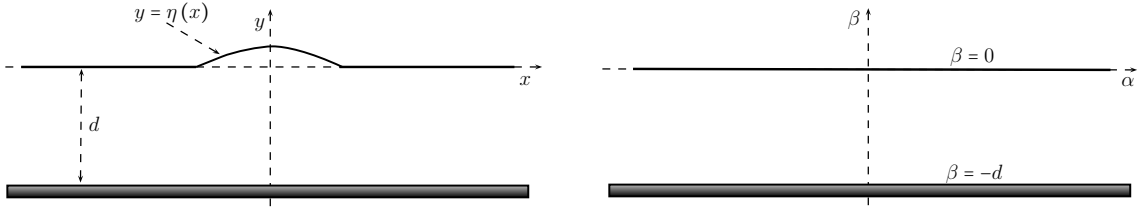
\begin{figure}
\centering
\scalebox{0.65} 
{
\begin{pspicture}(0,-2.0622187)(23.361876,2.0892189)
\definecolor{color88g}{rgb}{0.6941176470588235,0.6941176470588235,0.6941176470588235}
\definecolor{color88f}{rgb}{0.00392156862745098,0.00392156862745098,0.00392156862745098}
\psline[linewidth=0.025999999cm,linestyle=dashed,dash=0.16cm 0.16cm,arrowsize=0.05291667cm 2.0,arrowlength=1.4,arrowinset=0.4]{->}(0.0,0.79078126)(10.86,0.79078126)
\psline[linewidth=0.025999999cm,linestyle=dashed,dash=0.16cm 0.16cm,arrowsize=0.05291667cm 2.0,arrowlength=1.4,arrowinset=0.4]{->}(12.08,0.75078124)(22.9,0.73078126)
\psline[linewidth=0.025999999cm,linestyle=dashed,dash=0.16cm 0.16cm,arrowsize=0.05291667cm 2.0,arrowlength=1.4,arrowinset=0.4]{<-}(5.36,1.9707812)(5.36,-2.0492187)
\psline[linewidth=0.025999999cm,linestyle=dashed,dash=0.16cm 0.16cm,arrowsize=0.05291667cm 2.0,arrowlength=1.4,arrowinset=0.4]{<-}(17.46,1.9907813)(17.46,-2.0292187)
\psframe[linewidth=0.04,dimen=outer,fillstyle=gradient,gradlines=2000,gradbegin=color88g,gradend=color88f,gradmidpoint=1.0](10.6,-1.6092187)(0.02,-1.8892188)
\psframe[linewidth=0.04,dimen=outer,fillstyle=gradient,gradlines=2000,gradbegin=color88g,gradend=color88f,gradmidpoint=1.0](22.78,-1.5892187)(12.2,-1.8692187)
\psline[linewidth=0.05cm](12.7,0.75078124)(22.36,0.73078126)
\usefont{T1}{ptm}{m}{n}
\rput(10.521406,0.5157812){$x$}
\usefont{T1}{ptm}{m}{n}
\rput(5.1514063,1.6557813){$y$}
\psline[linewidth=0.025999999cm,linestyle=dashed,dash=0.16cm 0.16cm,arrowsize=0.05291667cm 2.0,arrowlength=1.4,arrowinset=0.4]{<-}(4.34,1.0507812)(3.26,1.6907812)
\usefont{T1}{ptm}{m}{n}
\rput(3.3714063,1.8957813){$y = \eta\,(x)$}
\usefont{T1}{ptm}{m}{n}
\rput(22.711407,0.49578124){$\alpha$}
\usefont{T1}{ptm}{m}{n}
\rput(17.171406,1.7557813){$\beta$}
\psline[linewidth=0.025999999cm,linestyle=dashed,dash=0.16cm 0.16cm,arrowsize=0.05291667cm 2.0,arrowlength=1.4,arrowinset=0.4]{<->}(1.54,0.75078124)(1.52,-1.5692188)
\usefont{T1}{ptm}{m}{n}
\rput(1.7914063,-0.36421874){$d$}
\usefont{T1}{ptm}{m}{n}
\rput(19.601406,-1.3842187){$\beta = -d$}
\usefont{T1}{ptm}{m}{n}
\rput(19.561407,0.99578124){$\beta = 0$}
\psline[linewidth=0.05cm](0.32,0.79078126)(3.86,0.79078126)
\psline[linewidth=0.05cm](6.76,0.79078126)(10.3,0.79078126)
\pscustom[linewidth=0.04]
{
\newpath
\moveto(3.86,0.79078126)
\lineto(4.38,1.0007813)
\curveto(4.64,1.1057812)(5.155,1.2107812)(5.41,1.2107812)
\curveto(5.665,1.2107812)(6.145,1.1057812)(6.82,0.79078126)
}
\end{pspicture}
}
\caption{\small\em Sketch of physical and transformed domains.}
\label{fig:sketch}
\end{figure}

Consider the change of independent variable $z \mapsto \zeta \equiv (\ui\sur{\psi} - f)/c$, that conformally maps the fluid domain $\{-\infty\leqslant x\leqslant\infty;-\depth\leqslant y\leqslant\eta\}$ into the strip $\{-\infty \leqslant \alpha \leqslant \infty; -\depth \leqslant \beta \leqslant 0\}$ where $\alpha\equiv \Re(\zeta)$ and $\beta\equiv\Im(\zeta)$ (c.f. Fig. \ref{fig:sketch}). The conformal mapping yields the Cauchy--Riemann relations $x_\alpha = y_\beta$ and $x_\beta = -y_\alpha$, while the complex velocity and the velocity components are
\[
w\ =\ -\,c\left(\frac{\ud\,z}{\ud\/\zeta}\right)^{\!-1}, \qquad
u\ =\ \frac{-\,c\,x_\alpha}{x_\alpha^{\,2}\/+\/y_\alpha^{\,2}}, \qquad
v\ =\ \frac{-\,c\,y_\alpha}{x_\alpha^{\,2}\/+\/y_\alpha^{\,2}}, \qquad
u^2\,+\,v^2\ =\ \frac{c^2}{x_\alpha^{\,2}\/+\/y_\alpha^{\,2}}. 
\]

Introducing new dependent variables $X(\alpha,\beta)\equiv x-\alpha$ and $Y(\alpha,\beta)\equiv y-\beta$, the Cauchy--Riemann relations $X_\alpha=Y_\beta$ and $X_\beta=-Y_\alpha$ hold, while the bottom ($\beta=-\depth$) and the free surface ($\beta=0$) impermeabilities yield $\bot{Y}(\alpha) \equiv Y(\alpha,-\depth) = 0$ and $\sur{Y}(\alpha)\equiv Y(\alpha,0)=\eta$. The functions $X$ and $Y$ can be expressed in term of $\bot{X}$ solely using some pseudo-differential operators \cite{Clamond1999}
\begin{align}\label{eq:solyxbot}
X(\alpha,\beta)\ =\ \cos\!\left[\,(\beta + \depth)\/\partial_\alpha\,\right] \bot{X}(\alpha), \qquad
Y(\alpha,\beta)\ =\ \sin\!\left[\,(\beta+\depth)\/\partial_\alpha\,\right] \bot{X}(\alpha),
\end{align}
so that the Cauchy--Riemann relations and the bottom impermeability are fulfilled identically. At the free surface $\beta=0$, \eqref{eq:solyxbot} yields $\sur{X}(\alpha) = \cos\!\left[\depth\partial_\alpha \right]\bot{X}(\alpha)$, that can be inverted as $\bot{X}(\alpha)= \sec\!\left[\depth\partial_\alpha \right]\sur{X}(\alpha)$, and hence the relation \eqref{eq:solyxbot} yields 
\begin{equation}\label{eq:relYXs}
\sur{Y}(\alpha)\ =\ \eta(\alpha)\ =\ \tan\!\left[\,\depth\,\partial_\alpha\,\right]\sur{X}(\alpha),
\end{equation}
which relates quantities written at the free surface only. The relation \eqref{eq:relYXs} can be trivially inverted giving, in particular, $\sur{(\partial_\alpha X)} = \mathscr{C}\{\sur{Y}\}\equiv 
\partial_\alpha \cot\!\left[\/\depth\/\partial_\alpha\/\right]\sur{Y}$, where $\mathscr{C}$ is a self-adjoint pseudo-differential operator acting on a pure frequency as
\begin{equation}\label{eq:cop}
\mathscr{C}\left\{\ue^{\ui k\alpha}\right\}\,=\,\left\{
\begin{array}{lr}
k\coth(k\depth)\,
\ue^{\ui k\alpha} &\quad (k\neq0)  \\
1/\depth & \quad  (k=0)
\end{array}
\right.
\end{equation}
This operator can be efficiently evaluated in the Fourier space using a FFT algorithm.

\subsection{Babenko's equation}

A Lagrangian for water waves can be obtained considering the kinetic energy minus the potential one, leading, in the conformal plane, to the functional $\L= \int_{-\infty}^{\infty}\,\mathfrak{L} \, \ud\/\alpha$ with $\Lf$ being the Lagrangian density \cite{Okasho2002}
\begin{equation*}
\Lf\ =\ \half\,c^2\,\eta\,\mathscr{C}\{\eta\}\ -\ \half\,g\,\eta^2\,
(\/1\,+\,\mathscr{C}\{\eta\}\/).
\end{equation*}
The governing equation is the Euler--Lagrange equation of this functional, i.e.
\begin{equation} \label{bablag}
0\ =\ \frac{\partial\,\mathfrak{L}}{\partial\/\eta}\ +\ \mathscr{C}\!\left\{\frac{\partial\,
\mathfrak{L}}{\partial\,\mathscr{C}\{\eta\}}\right\}\ =\ c^2\,\mathscr{C}\{\eta\}\ -\ g\,\eta\ 
-\ \half\, g\,\mathscr{C}\{\eta^2\}\ -\ g\,\eta\,\mathscr{C}\{\eta\},
\end{equation}
which is the Babenko equation for gravity solitary surface waves \cite{Babenko1987}. Applying the operator $\mathscr{C}^{-1}$ to (\ref{bablag}) and splitting the linear and nonlinear parts, the equation can be presented as 
\begin{equation}\label{eq:babpet}
\Ll\{\eta\}\, =\, \Nn\{\eta\}, \quad
\Ll\{\eta\}\, \equiv\, c^2\,\eta\, -\, g\,\mathscr{C}^{-1}\{\eta\}, \quad
\Nn\{\eta\}\, \equiv\, g\,\mathscr{C}^{-1}\!\left\{\/\eta\,\mathscr{C}\{\eta\}
\/\right\}\, +\, \half\,g\,\eta^2,
\end{equation}
that is more convenient for the simple and efficient numerical resolution explained below.

\section{Numerical method and results}\label{sec:num}

First, we define the wave by choosing its Froude number $F\equiv c/\sqrt{gd}$. Second, we choose a KdV-like initial guess of the free surface elevation $\eta_{0} (\alpha)$
\begin{equation}\label{eq:serre}
\eta_0(\alpha)\ \equiv\ d\,(F^2-1)\,\sech^2(\kappa\/\alpha/2), \quad (a) \qquad F^2\ =\ \tan(\kappa\/d)\,/\,\kappa\/d \quad (b).
\end{equation}
Third, equation (\ref{eq:serre}-{\it b}) is used to compute $\kappa$ and thence to determine the length $L$ of the computational domain (i.e. $-L \leqslant \alpha \leqslant L$) such that $\eta_0(L) = 0$ to machine precision. The computational domain is then discretised with $N$ equally spaced nodes, with $N$ large enough to avoid aliasing. Fourth, equation \eqref{eq:babpet} is solved via Petviashvili's iterations \cite{Lakoba2007, Pelinovsky2004, Petviashvili1976}
\begin{equation}\label{eq:pet2}
\eta_{n+1}\ =\ S_n^{\,2}\,\/\Ll^{-1}\left\{\Nn\{\eta_n\}\right\}, \qquad
S_n\ =\ \frac{\int_{-\infty}^\infty{\eta}_n\,\Ll\{\eta_n\}\,
\ud\/\alpha}{\int_{-\infty}^\infty{\eta}_n\,\Nn\{\eta_n\}\,
\ud\/\alpha},
\end{equation}
where $S_n$ is the co-called {\em stabilisation factor\/} which can be computed either in physical or Fourier spaces, thanks to the Parseval identity \cite{Titchmarsh1976}. In the implementation, we systematically privilege the Fourier space since the operators $\mathscr{C}$ and $\mathscr{C}^{-1}$ can be very efficiently computed according to the definition \eqref{eq:cop} using any FFT algorithm.

The convergence of the iterative process \eqref{eq:pet2} is checked by following the norm of the difference between two successive iterations along with the residual in the $\ell_\infty$ norm
\begin{equation*}
\|\,\eta_{n+1}\, -\, \eta_{n}\, \|_\infty\ <\ \eps_1, \qquad
\|\,\Ll\{\eta_n\}\, -\, \Nn\{\eta_n\}\,\|_\infty\ <\ \eps_2,
\end{equation*}
where $\eps_n$ are some prescribed tolerance parameters that are usually of the order of the floating point arithmetics precision.

\subsection{Numerical results}

The present algorithm is illustrated on the following example of a fairly high amplitude solitary wave. It is validated and compared against Tanaka's algorithm \cite{Tanaka1986}. Tanaka's solution is parametrized by the dimensionless parameter $q_c = |u_c|/\sqrt{gd} \in (0,1)$, where $u_c$ is the fluid velocity at the wave crest. We choose the special value $q_c = 0.4$ for comparisons. Tanaka's algorithm was implemented in {\sc Matlab}\footnote{In the sequel, all the algorithms are compared in the same computing environment: {\sc Matlab} \textsuperscript{\textregistered}.} in its original version without any peculiar optimisations. The interval $-30\leqslant\phi\leqslant30$ is discretised using $20,001$ nodes and the iterations are continued until the tolerance $10^{-10}$ is obtained between two successive values of the Froude number $F$, i.e.
\begin{equation*}
  |\,F_{n+1}^{\,2}\, -\, F_n^{\,2}\,|\ <\ \eps_3\ (=10^{-10} \ \ \mbox{in our computation).}
\end{equation*}
This computation required $47$ iterations which lasted $385\,\mathsf{s}$ on our desktop computer. The resulting Froude number computed by Tanaka's algorithm is 
\begin{equation*}
  F\ =\ 1.274236297977903.
\end{equation*}

Then, we take this Froude number and use it as the solitary wave parameter in the Babenko equation \eqref{eq:babpet}. We use a (periodised) domain of the same length as in Tanaka's method, and discretise it with $N = 16,384$ nodes. The tolerance is chosen to be $10^{-15}$, i.e. very close to machine accuracy. These parameters are kept in all the computations presented below. The iterative procedure is stopped when the $\ell_\infty$ norm of the difference between two successive iterations is less than the tolerance (the norm of the residual is checked a posteriori). Our algorithm required $372$ iterations to converge and the computations lasted for about $0.5\,\mathsf{s}$, i.e. almost three orders of magnitude faster than Tanaka's algorithm. This flagrant difference in CPU times can be explained by the fact that our method has the super-linear complexity $\O(N\log N)$ while Tanaka's method has a quadratic complexity $\O(N^2)$.

\begin{figure}
  \centering
  \subfigure[]%
  {\includegraphics[width=0.49\textwidth]{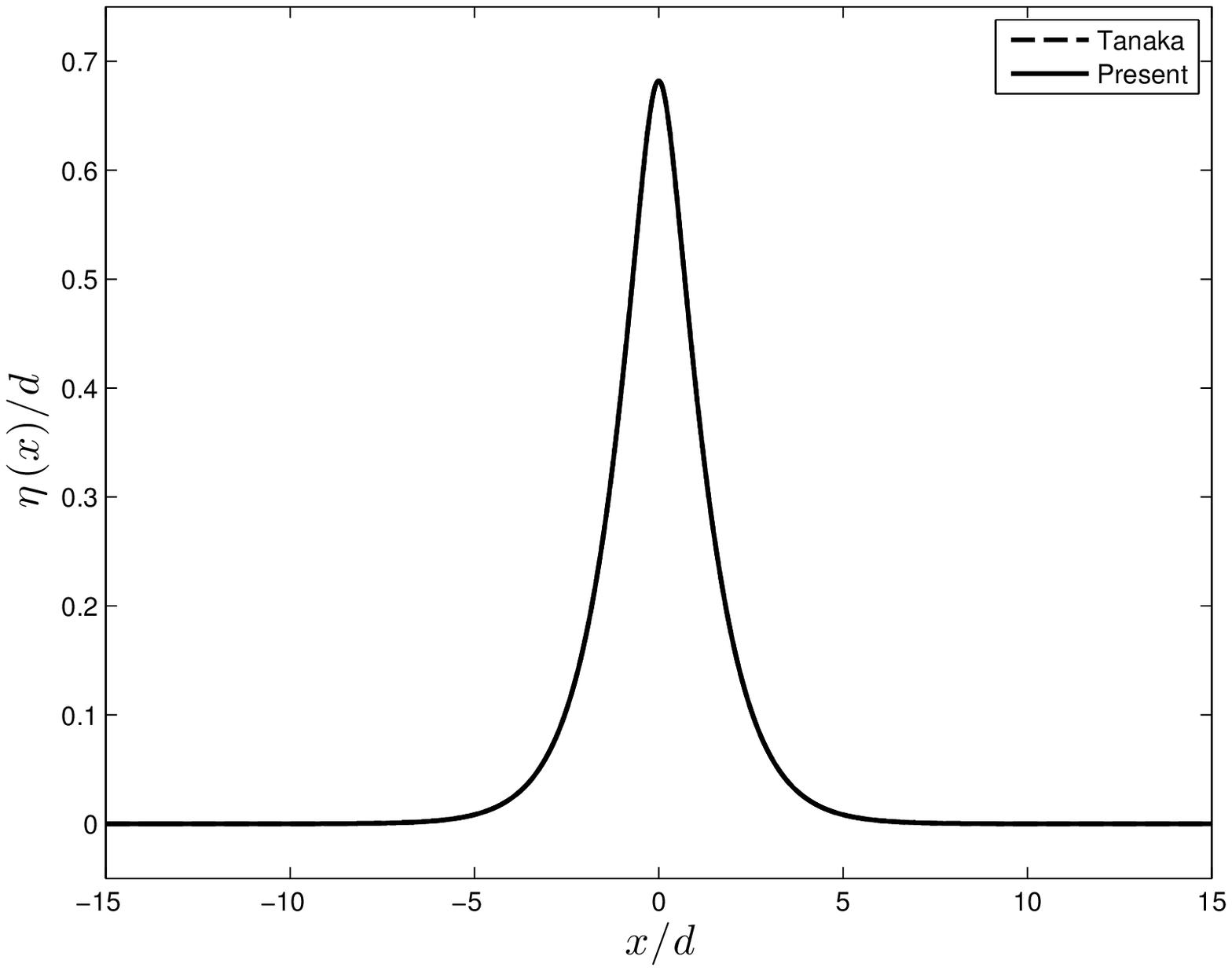}}
  \subfigure[]%
  {\includegraphics[width=0.49\textwidth]{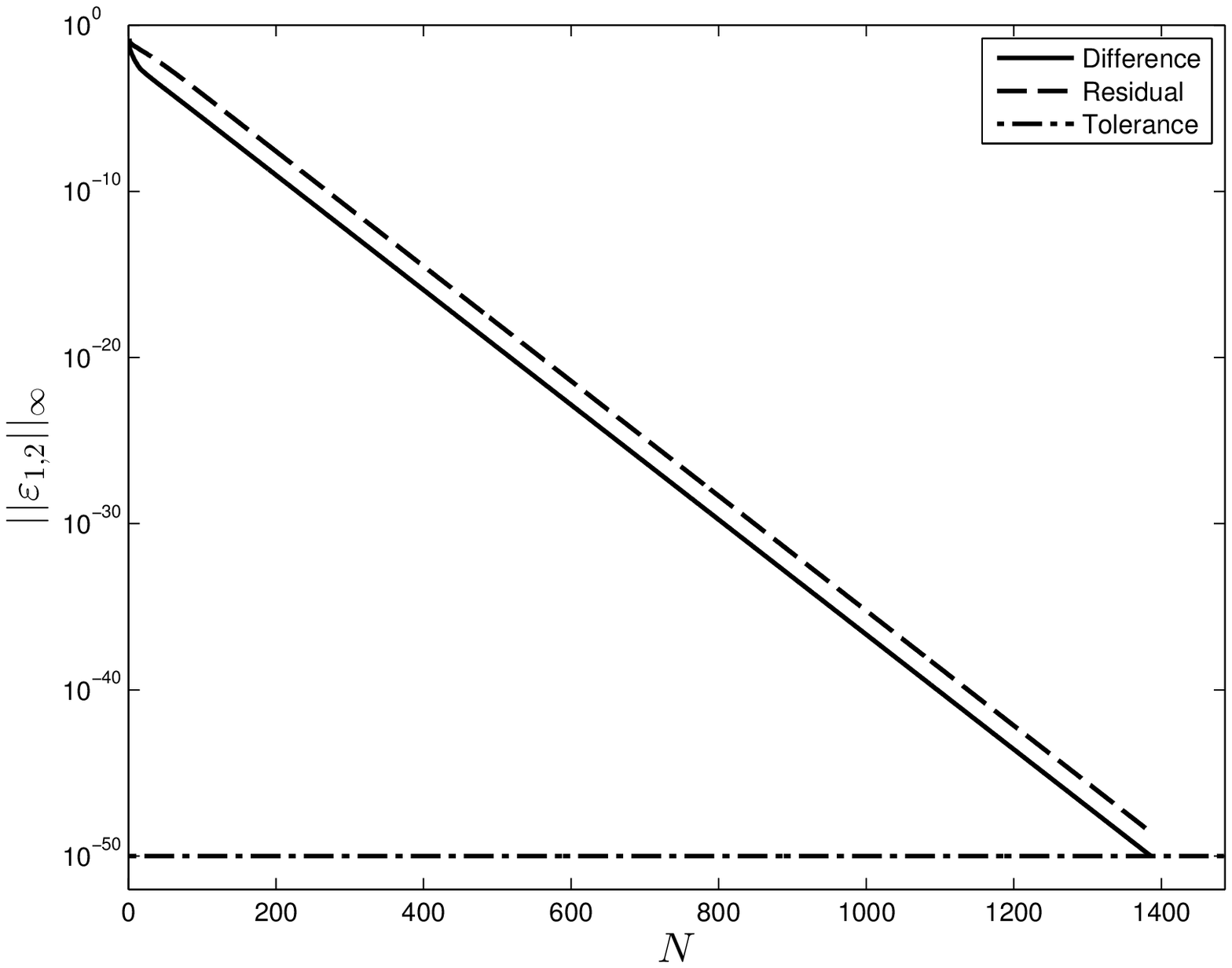}}
  \caption{\small\em Solitary wave profile (a) and convergence of the numerical error during 
  the iterations (b).}
  \label{fig:sws}
\end{figure}

Moreover, the present algorithm can be used to compute solitary waves to an arbitrary precision. To illustrate this, we implemented the scheme using a Multi-Precision Computing Toolbox (MPCT) for {\sc Matlab} \cite{MATLAB2012}. The code modifications needed to implement the arbitrary accuracy are minor; this constitutes one of the major advantages of this MPCT. In these higher-precision experiments, all floating point operations are done with $50$ significant digits. The tolerance parameter is set to $10^{-50}$. To achieve such extreme accuracy, the Petviashvili scheme required $1,386$ iterations. The convergence of the algorithm is shown on Figure~\ref{fig:sws}-{b}. One can notice that the residual error is slightly higher than the difference between two successive iterations, likely due to the presence of quadratic nonlinearities into the equation. However, both measures of the numerical error decrease monotonically with iterations. The resulting solitary wave profile is depicted on Figure~\ref{fig:sws}-a along with the corresponding Tanaka solution. The two curves cannot be distinguished up to the graphical resolution, thus validating our method. The respective accuracies can be asserted comparing the dimensionless amplitudes $a/d$ (Table~\ref{table1}).

\begin{table}
\scriptsize
\centering
\begin{tabular}{ccc}
Tanaka (16 digits) & Present (16 digits) & Present (50 digits) \\
\hline
0.6819\underline{448200954614} & 0.6819656673869\underline{021} &
0.681965667386915720709857779975770373038440370273\underline{43} 
\end{tabular}
\vspace{2mm}
\caption{\em\small Amplitude corresponding to $F=1.274236297977903$ (wrong digits 
are underlined).} \label{table1}
\end{table}

One can notice that despite the tolerance parameter set to $10^{-10}$ in Tanaka's algorithm, its real accuracy is seriously degraded for this large amplitude wave (only five digits of the wave amplitude are predicted correctly). Conversely, the present method gives fourteen correct digits with a tolerance of $10^{-15}$ in double precision calculation (about sixteen digits), and 49 correct digits using the 50-digit computation (the latter was determined using a 60-digit calculation with a tolerance of $10^{-55}$ with eight times more nodes over an interval twice the size).

\begin{figure}
  \centering
  \includegraphics[width=0.59\textwidth]{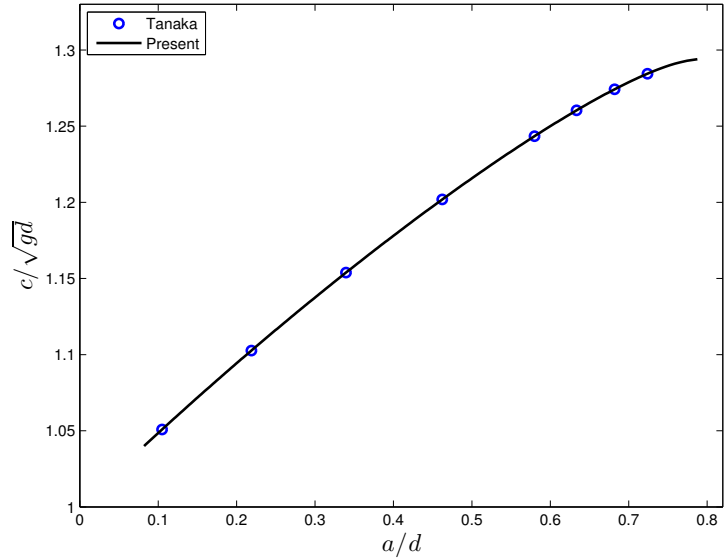}
  \caption{\small\em Speed versus amplitude relation predicted by the present (solid line) and 
  Tanaka's (circles) algorithms.}
  \label{fig:csa}
\end{figure}

The algorithm being validated, it can be used to produce some physically sound numerical results. For example, we compute to high-precision (here $10^{-15}$, but it is not a limitation) the speed--amplitude relation for solitary waves. The result is shown on Figure~\ref{fig:csa}. We also computed several values in a wide range of amplitudes using Tanaka's algorithm (represented with circles). Again, up to the graphical resolution we have a very good agreement between the two methods. However, the speed of Tanaka's algorithm does not allow to compute many values in a reasonable time.

\begin{figure}
  \centering
  \includegraphics[width=0.59\textwidth]{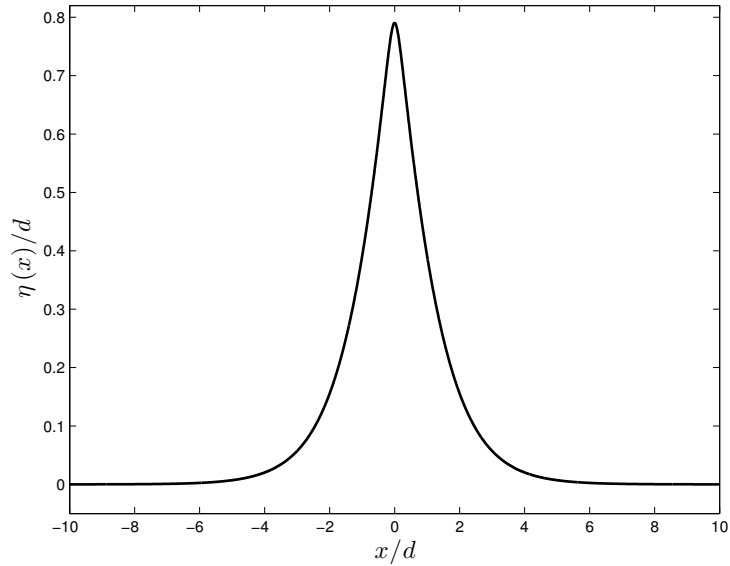}
  \caption{\small\em A very high amplitude solitary wave ($a/d \approx 0.790$).}
  \label{fig:high}
\end{figure}

We note finally that, for simplicity, we defined the wave by its Froude number which varies in the interval $1\leqslant F\lesssim 1.29421$. Doing so, we can compute all waves such that $0\leqslant a\lesssim 0.796$ that is sufficient for all cases of practical interest. One of the largest solitary waves which can be predicted by this parametrization is represented on Figure~\ref{fig:high}. However, the solutions of larger amplitudes exist, the limit $a\lesssim 0.796$ being obtained here because $F$ is not a monotonic function of $a$ (i.e., $F(a)$ is multi-valued when $a$ is large). Thus, in order to compute larger waves, another parameter (e.g., $a/d$ or $q_c$) should be used to define the wave. This requires simple modifications in the code \cite{Clamond2012}. We successfully did it and were able to compute all waves such that $a\lesssim 0.82$. For even higher waves, the present method is not efficient because these extreme waves being (almost) singular at the crest, the Fourier series does not converge fast enough and another spectral decomposition should be used instead. 

\section{Conclusions}\label{sec:concl}

We have described a fast, accurate and easy to implement method for computing steady gravity solitary surface waves solution of the full Euler equations. The method is based on first rewriting the problem into the equivalent but simpler Babenko's equation and, second, using Petviashvili's iterations to solve it numerically. The resulting scheme allows the computation to any arbitrary accuracy using a multi-precision floating-point arithmetics library \cite{MATLAB2012}. All ingredients of the method are well known, but their combination seems to be new in this context and turns out to be very efficient. Our {\sc Matlab} implementation is rather compact: The computational numerical core is about a dozen lines of code, the rest being pre- and post-processing. The script \cite{Clamond2012} can be freely downloaded and used (it can also be easily translated into any programming language). 

\section*{Acknowledgements}

Denys~\textsc{Dutykh} acknowledges the support from ERC under the research project ERC-2011-AdG 290562-MULTIWAVE. The authors would like to thank Pavel~\textsc{Holodoborodko} for invaluable help with the multi-precision floating point arithmetics.

\end{document}